\documentstyle[12pt]{article}

\skewchar\fivmi='177
\skewchar\sixmi='177
\skewchar\sevmi='177
\skewchar\egtmi='177
\skewchar\ninmi='177
\skewchar\tenmi='177
\skewchar\elvmi='177
\skewchar\twlmi='177
\skewchar\frtnmi='177
\skewchar\svtnmi='177
\skewchar\twtymi='177
\def\@magscale#1{ scaled \magstep #1}
\font\twfvmi  = ammi10   \@magscale5 
\skewchar\twfvmi='177
\skewchar\fivsy='60
\skewchar\sixsy='60
\skewchar\sevsy='60
\skewchar\egtsy='60
\skewchar\ninsy='60
\skewchar\tensy='60
\skewchar\elvsy='60
\skewchar\twlsy='60
\skewchar\frtnsy='60
\skewchar\svtnsy='60
\skewchar\twtysy='60
\font\twfvsy  = amsy10   \@magscale5 
\skewchar\twfvsy='60

\catcode`@=11
\def\un#1{\relax\ifmmode\@@underline#1\else
        $\@@underline{\hbox{#1}}$\relax\fi}
\catcode`@=12

\def\a{\alpha}
\def\b{\beta}

\def\d{\delta}
\def\e{\epsilon}

\def\g{\gamma}
\def\h{\eta}

\def\j{\psi}

\def\q{\theta}

\def\s{\sigma}
\def\t{\tau}

\def\F{\Phi}
\def\G{\Gamma}
\def\J{\Psi}
\def\L{\Lambda}
\def\O{\Omega}

\def\S{\Sigma}

\def\cb{{\cal B}}

\def\cv{{\cal V}}

\font\sc=font005                        
\font\ooo=circle10                      
\font\ro=manfnt                         
\def\kcl{{\hbox{\ro 6}}}                
\def\kcr{{\hbox{\ro 7}}}                
\def\ktl{{\hbox{\ro \char'134}}}        
\def\ktr{{\hbox{\ro \char'135}}}        
\def\kbl{{\hbox{\ro \char'136}}}        
\def\kbr{{\hbox{\ro \char'137}}}        

\def\bo{{\raise.15ex\hbox{\large$\Box$}}}               
\def\pa{\partial}                                       
\def\TH{{\raise.2ex\hbox{$\displaystyle \bigodot$}\mskip-4.7mu \llap H \;}}
\def\face{{\raise.2ex\hbox{$\displaystyle \bigodot$}\mskip-2.2mu \llap {$\ddot
        \smile$}}}                                      
\def\dg{\sp\dagger}                                     

\def\sp#1{{}^{#1}}                              
\def\Tilde#1{\widetilde{#1}}                    
\def\Hat#1{\widehat{#1}}                        
\def\Bar#1{\overline{#1}}                       
\def\leftrightarrowfill{$\mathsurround=0pt \mathord\leftarrow \mkern-6mu
        \cleaders\hbox{$\mkern-2mu \mathord- \mkern-2mu$}\hfill
        \mkern-6mu \mathord\rightarrow$}
\def\dvec#1{\vbox{\ialign{##\crcr
        \leftrightarrowfill\crcr\noalign{\kern-1pt\nointerlineskip}
        $\hfil\displaystyle{#1}\hfil$\crcr}}}           
\def\dt#1{{\buildrel {\hbox{\LARGE .}} \over {#1}}}     

\def\frac#1#2{{\textstyle{#1\over\vphantom2\smash{\raise.20ex
        \hbox{$\scriptstyle{#2}$}}}}}                   
\def\sfrac#1#2{{\vphantom1\smash{\lower.5ex\hbox{\small$#1$}}\over
        \vphantom1\smash{\raise.4ex\hbox{\small$#2$}}}} 
\def\bfrac#1#2{{\vphantom1\smash{\lower.5ex\hbox{$#1$}}\over
        \vphantom1\smash{\raise.3ex\hbox{$#2$}}}}       
\def\afrac#1#2{{\vphantom1\smash{\lower.5ex\hbox{$#1$}}\over#2}}    

\newskip\humongous \humongous=0pt plus 1000pt minus 1000pt
\def\caja{\mathsurround=0pt}
\def\eqalign#1{\,\vcenter{\openup2\jot \caja
        \ialign{\strut \hfil$\displaystyle{##}$&$
        \displaystyle{{}##}$\hfil\crcr#1\crcr}}\,}
\newif\ifdtup

\def\ref#1{$\sp{#1)}$}

\parskip=\medskipamount                 
\thicklines                         

\def\oldheadpic{                                
        \setlength{\unitlength}{.4mm}
        \thinlines
        \par
        \begin{picture}(349,16)
        \put(325,16){\line(1,0){4}}
        \put(330,16){\line(1,0){4}}
        \put(340,16){\line(1,0){4}}
        \put(335,0){\line(1,0){4}}
        \put(340,0){\line(1,0){4}}
        \put(345,0){\line(1,0){4}}
        \put(329,0){\line(0,1){16}}
        \put(330,0){\line(0,1){16}}
        \put(339,0){\line(0,1){16}}
        \put(340,0){\line(0,1){16}}
        \put(344,0){\line(0,1){16}}
        \put(345,0){\line(0,1){16}}
        \put(329,16){\oval(8,32)[bl]}
        \put(330,16){\oval(8,32)[br]}
        \put(339,0){\oval(8,32)[tl]}
        \put(345,0){\oval(8,32)[tr]}
        \end{picture}
        \par
        \thicklines
        \vskip.2in}
\def\oldtitle#1#2#3#4{\oldheadpic\begin{center}\vglue.5in{\large\bf #1}\\[.6in]
        {#2}\\[.1in] {\it Department of Physics and Astronomy}\\
        {\it University of Maryland, College Park, MD 20742}\\[.6in]
        Physics Publication \#{#3}\\ {#4}\\[1.5in] {\bf ABSTRACT}\\[.1in]
        \end{center} \begin{quotation}}                 
\def\oldTitle#1#2#3#4#5#6#7{\oldheadpic\begin{center} \vglue .4in
        {\large\bf #1}\\[.4in]
        {#2}\\[.1in] {\it Department of Physics and Astronomy}\\
        {\it University of Maryland, College Park, MD 20742}\\[.1in]
        {#3}\\[.1in] {\it {#4}}\\ {\it {#5}}\\[.4in]
        Physics Publication \#{#6}\\ {#7}\\[.5in] {\bf ABSTRACT}\\[.1in]
        \end{center} \begin{quotation}}                 
\def\border{                                            
        \setlength{\unitlength}{1mm}
        \newcount\xco
        \newcount\yco
        \xco=-24
        \yco=12
        \begin{picture}(140,0)
        \put(\xco,\yco){$\ktl$}
        \advance\yco by-1
        {\loop
        \put(\xco,\yco){$\kcl$}
        \advance\yco by-2
        \ifnum\yco>-240
        \repeat
        \put(\xco,\yco){$\kbl$}}
        \xco=158
        \yco=12
        \put(\xco,\yco){$\ktr$}
        \advance\yco by-1
        {\loop
        \put(\xco,\yco){$\kcr$}
        \advance\yco by-2
        \ifnum\yco>-240
        \repeat
        \put(\xco,\yco){$\kbr$}}
        \put(-20,11){\tiny University of Maryland Elementary Particle
Physics University of Maryland Elementary Particle Physics University of
Maryland Elementary Particle Physics}
        \put(-20,-241.5){\tiny University of Maryland Elementary
Particle Physics University of Maryland Elementary Particle Physics
University of Maryland Elementary Particle Physics}
        \end{picture}
        \par\vskip-8mm}
\def\bordero{                                           
        \setlength{\unitlength}{1mm}
        \newcount\xco
        \newcount\yco
        \xco=-24
        \yco=12
        \begin{picture}(140,0)
        \put(\xco,\yco){$\ktl$}
        \advance\yco by-1
        {\loop
        \put(\xco,\yco){$\kcl$}
        \advance\yco by-2
        \ifnum\yco>-240
        \repeat
        \put(\xco,\yco){$\kbl$}}
        \xco=158
        \yco=12
        \put(\xco,\yco){$\ktr$}
        \advance\yco by-1
        {\loop
        \put(\xco,\yco){$\kcr$}
        \advance\yco by-2
        \ifnum\yco>-240
        \repeat
        \put(\xco,\yco){$\kbr$}}
        \put(-20,12){\ooo
bacdefghidfghghdhededbihdgdfdfhhdheidhdhebaaahjhhdahbahgdedgehgfdiehhgdigicba}
        \put(-20,-241.5)
{\ooo
ababaighefdbfghgeahgdfgafagihdidihiidhiagfedhadbfdecdcdf
agdcbhaddhbgfchbgfdacfediacbabab}
        \end{picture}
        \par\vskip-8mm}
\def\headpic{                                           
        \indent
        \setlength{\unitlength}{.4mm}
        \thinlines
        \par
        \begin{picture}(29,16)
        \put(165,16){\line(1,0){4}}
        \put(170,16){\line(1,0){4}}
        \put(180,16){\line(1,0){4}}
        \put(175,0){\line(1,0){4}}
        \put(180,0){\line(1,0){4}}
        \put(185,0){\line(1,0){4}}
        \put(169,0){\line(0,1){16}}
        \put(170,0){\line(0,1){16}}
        \put(179,0){\line(0,1){16}}
        \put(180,0){\line(0,1){16}}
        \put(184,0){\line(0,1){16}}
        \put(185,0){\line(0,1){16}}
        \put(169,16){\oval(8,32)[bl]}
        \put(170,16){\oval(8,32)[br]}
        \put(179,0){\oval(8,32)[tl]}
        \put(185,0){\oval(8,32)[tr]}
        \end{picture}
        \par\vskip-6.5mm
        \thicklines}
\def\title#1#2#3#4{\border\headpic {\hbox to\hsize{#4 \hfill UMDEPP #3}}\par
        \begin{center} \vglue .5in {\large\bf #1}\\[.6in]
        {#2}\\[.1in] {\it Department of Physics and Astronomy}\\
        {\it University of Maryland, College Park, MD 20742}\\[1.5in]
        {\bf ABSTRACT}\\[.1in] \end{center} \begin{quotation}}  
\def\Title#1#2#3#4#5#6#7{\border\headpic
        {\hbox to\hsize{#7 \hfill UMDEPP #6}}\par
        \begin{center} \vglue .4in {\large\bf #1}\\[.4in]
        {#2}\\[.1in] {\it Department of Physics and Astronomy}\\
        {\it University of Maryland, College Park, MD 20742}\\[.1in]
        {#3}\\[.1in] {\it {#4}}\\ {\it {#5}}\\[.5in] {\bf ABSTRACT}\\[.1in]
        \end{center} \begin{quotation}}                 
\def\endtitle{\end{quotation}\newpage}                  

\begin{document}

\def\[{\lfloor{\hskip 0.35pt}\!\!\!\lceil}
\def\]{\rfloor{\hskip 0.35pt}\!\!\!\rceil}
\def\ve{\varepsilon}

\def\ul{\underline}
\def\ula{{\underline a}}
\def\ulb{{\underline b}}
\def\ulc{{\underline c}}
\def\uld{{\underline d}}

\def\ab{\bar{a}}
\def\bb{\bar{b}}
\def\cb{\bar{c}}
\def\db{\bar{d}}

\def\st{\Tilde{\s}}
\def\3b{\bar{3}}

\def\dta{\dt{\a}}
\def\dtb{\dt{\b}}

\def\gfrac#1#2{\frac {\scriptstyle{#1}}
        {\mbox{\raisebox{-.6ex}{$\scriptstyle{#2}$}}}}
\def\gg{{\hbox{\sc g}}}
\border\headpic {\hbox to\hsize{
February 1992 \hfill {UMDEPP 92-163}}}\par
\begin{center}
\vglue .4in
{\large\bf Majorana-Weyl Spinors and Self-Dual Gauge Fields \\
in $2 + 2$ Dimensions}
\\[.2in]
Sergei V. KETOV\footnote{On leave of absence from: High Current Electronics
Institute of the Russian Academy of Sciences, Siberian Branch,
 Akademichesky~4, Tomsk 634055, Russia}\\[.1in]
{\it Department of Physics \\
University of Maryland at College Park\\
College Park, MD 20742-4111 USA}
\\[.1in]
Hitoshi NISHINO and S. James GATES Jr.\footnote{The research of H. Nishino and
S. J. Gates Jr. was supported by NSF grant \# PHY--91--19746}\\[.1in]
{\it Department of Physics \\
University of Maryland at College Park\\
College Park, MD 20742-4111 USA}\\[.1in]
and
\\[.1in]
{\it Department of Physics and Astronomy \\
Howard University \\
Washington, D.C. 20059, USA}\\
[.2in]

{\bf ABSTRACT}\\[.1in]
\end{center}
\begin{quotation}

The properties of Dirac gamma matrices in a four-dimensional space-time with
the $(2,2)$ signature  are studied. The basic spinors are classified, and the
existence of Majorana-Weyl spinors is noted. Supersymmetry in $2 + 2$
dimensions is discussed, and the existence of the {\it real chiral} scalar
supermultiplet is discovered. Supersymmetric {\it self-dual} Yang-Mills
theories and {\it self-dual} supergravity model in $2 + 2$ dimensions, that
are apparently relevant to integrable systems, are formulated for the first
time.

\endtitle

\newpage

\oddsidemargin=0.03in
\evensidemargin=0.01in
\hsize=6.5in
\textwidth=6.5in

\noindent{\bf 1.~~Introduction}

The idea to consider space-time with more than one {\it time} component is
very old. Obviously, such a space-time does not correspond to our observed
universe, and the lack of much interest in this possibility in the past had
at least two-fold reasoning: (i) an indefinite signature of space-time is
apparently unphysical, and (ii) there is no natural motivation to consider
space-time with such signature. Nevertheless, spaces with several time
components and indefinite signature do appear among the solutions of some
physically interesting systems in higher dimensions, and they have been studied
indeed in the compactification contexts of Kaluza-Klein and string theories
[1].

We are not going to address in this paper the first issue as to how to make
physical sense of multiple time theory. One may imagine an ambitious
scenario of space-time with dynamical signature or a ``spontaneous symmetry
breaking'' mechanism of changing the signature but, clearly, it would be too
faraway. Instead, let us concentrate on the second issue and ask
whether there is a natural motivation for introducing a space-time with more
than one temporal direction. By a ``natural motivation'' we mean the existence
of some kind of ``fundamental'' or ``master'' theory which would {\it require}
 the indefinite signature. Looking around, one easily finds that there is such
theory! It has been known since 1976 that there are strings with $N=2$ extended
{\it local} superconformal symmetry on a world-sheet which live in 2
space-time dimensions [2]. Later on, it was realized [3] that these are in fact
2 {\it complex} dimensions, and the correct signature is $(2,2)$. Most
recently, it was discovered by Ooguri and Vafa in a series of papers [4] that
the consistent backgrounds for $N=2$ string propagation correspond to {\it
self-dual} gravity configurations in the case of {\it closed} $N=2$ strings,
and {\it self-dual} Yang-Mills configurations (coupled to gravity) in the
case of $N=2$ {\it heterotic} strings, in four or lower dimensions. The result
of Ooguri and Vafa, as for the Yang-Mills sector of the $N=2$ heterotic string
is concerned, has recently been confirmed by Nishino and Gates [5] on the basis
of the $\b$-function calculations. Taking into account that the self-dual
Yang-Mills theory in four dimensions with the signature $(2,2)$ [6] is likely
to be considered as the generating model for all {\it integrable} systems in
dimensions lower than four [7], it seems to be enough motivation to study field
theories formulated in a four-dimensional space-time with a {\it complex}
time, and, especially, those among them which are supersymmetric and/or
self-dual. Surprisingly enough, a lot of issues related to such theories has
never been addressed, to our knowledge.

In this Letter we begin with the analysis of {\it spinors} living in $2 + 2$
dimensions. Then we apply it to describe the supersymmetric  self-dual
Yang-Mills and supergravity theories in space-time with the $(2,2)$ signature.
The self-dual super-Yang-Mills--supergravity system as a background of the
Green-Schwarz superstring will be discussed separately [8].

\newpage

\noindent{\bf 2.~~Dirac matrices}

Given flat space-time with four dimensions and the signature $(2,2)$, the
fundamental metric form $ds^2=\h_{ij}dy^idy^j$ in {\it real} coordinates $y^j =
 \left( y^1,y^2,y^3,y^4\right)$ and with the flat metric $\h_{ij}={\rm diag}(+,
+,-,-)$, can be rewritten in {\it complex} notation as
$$ds^2\equiv \h_{\ula\ulb}dx^{\ula}dx^{\ulb}=2\left(dx^1d\bar{x}^1 -dx^2d
\bar{x}^2\right) \equiv 2\h_{a\bb}dx^{a}dx^{\bb}~,\eqno(2.1) $$
where we have introduced the complex coordinates
$$x^1 = {1\over\sqrt{2}}\left( y^1 + iy^2\right)~,~~~x^2 = {1\over\sqrt{2}}
\left( y^3 + iy^4\right) \eqno(2.2) $$
and their conjugates, $x^i = \left( x^1,x^2,\bar{x}^1,\bar{x}^2\right)$,
instead of the real ones, and the notation
$$x^{a}=(x^1,x^2),~x^{\ab} = \bar{x}^{a}=(\bar{x}^1,\bar{x}^2);~~
\ula = (a,\ab),~a = 1,2,~\ab = \bar{1},\bar{2}~. \eqno(2.3) $$
The complex notation is natural in the $(2,2)$ space-time since
its structure group can equally be represented by either $SO(2,2)$ or $U(1,1)$.
In other words, there is an obvious complex structure in this space-time which
gives the simplest example of a four-dimensional hermitian manifold. The
reduced metric $\h_{a\bb}$ in eq.(2.1) takes the form
$$\h_{a\bb} =\h_{\bb a}= \left( \begin{array}{cc} 1 & 0 \\ 0 & -1 \end{array}
\right) ~. \eqno(2.4) $$
The inner product of $(2,2)$ vectors $V$ and $U$ reads: $V^{\ula}
U_{\ula}\equiv V^a U_a + V^{\ab}U_{\ab}=V^a\h_{a\bb}U^{\bb} + V^{\ab}\h_{\ab b}
U^b$. Thus the {\it Klein-Gordon} equation for a scalar $\F$ in the real
coordinates,
$$\left( \bo - m^2 \right)\F (y) \equiv \left( \h^{ij}\pa_{i}\pa_{j}
- m^2\right)\F (y) = 0~,\eqno(2.5) $$
is rewritten in the complex coordinates as
$$\left( \bo - m^2 \right) \F (x) \equiv \left( 2\h^{a\bb}\pa_{a}\pa_{\bb} -
m^2 \right) \F (x) = 0~.\eqno(2.6) $$

The {\it Dirac equation} results from the factorization procedure of the
Klein-Gordon kinetic operator in eq.(2.5), and  it takes the form
$$\left( i\G^j \pa_j + m \right) \J (y) = 0~,\eqno(2.7) $$
where the Dirac matrices $\G^j$ satisfy the Clifford algebra
$$ \left\{ \G^i,\G^j\right\} = 2\h^{ij}~.\eqno(2.8) $$
Eq.(2.6) can also be factorized, and it leads to the Dirac equation in the form
$$ \left(i\g^{\ula}\pa_{\ula} + m\right)\J (x) \equiv \left(i\g^{a}\pa_{a} +
i\g^{\ab}\pa_{\ab} + m \right) \J (x) = 0~,\eqno(2.9) $$
where we have introduced the corresponding gamma matrices $\g^{\ula}=\left(
\g^{a},\g^{\ab}\right)$. The latter satisfy an algebra
$$\g^{a}\g^{b} + \g^{b}\g^{a} = 0~,~~ \g^{\ab}\g^{\bb} + \g^{\bb}\g^{\ab} = 0~,
{}~~\g^{a}\g^{\bb} + \g^{\bb}\g^{a} = 2\h^{a\bb}~.\eqno(2.10) $$

The representation theory of the Clifford algebra (2.8) can be developed along
the lines of the familiar $(3+1)$-dimensional case. There exists only one
non-trivial $4\times 4$ matrix representation of eq.(2.8), and its
(equivalent) explicit forms can easily be constructed by the use of $2\times 2$
Pauli matrices, which satisfy the algebra $\t_1\t_2=i\t_3\;$.

The {\it Majorana} representation of the $\G$-matrices takes the form
$$\eqalign{
\G^1 =\left(\begin{array}{cc} 0 & -i\t_3 \\ i\t_3 & 0\end{array}\right)~, &
{}~~\G^2 =\left(\begin{array}{cc} 0 & -i\t_1 \\ i\t_1 &
0\end{array}\right)~,\cr
\G^3 =\left(\begin{array}{cc} 0 & \t_2 \\ -\t_2 & 0\end{array}\right)~, &
{}~~\G^4 =\left(\begin{array}{cc} iI_2 & 0 \\ 0 & -iI_2\end{array}\right)~,}
\eqno(2.11) $$
where $0$ and $I_2$ are the $2\times 2$ zero and identity matrices,
respectively. In this representation all of the components of the
$\G$-matrices are {\it pure imaginary}, i.e. all the
entries of the $i\G^j$ in the Dirac equation are purely {\it real}.

Another explicit representation, which is particularly useful in supersymmetry,
is that in which the $\G_5$-matrix is {\it diagonal}. It provides a preferred
basis for introducing the 2-component notation for spinors in four space-time
dimensions [9], and generically has the form
$$\G^j = \left(\begin{array}{cc} 0 & \s^j \\ \st^j & 0 \end{array}\right)
\eqno(2.12) $$
with some $2\times 2$-dimensional entries $\s^j$ and $\st^j$. The appropriate
explicit representation is given by
$$\eqalign{
\G^1 = \left(\begin{array}{cc} 0 & -i\t_1 \\ +i\t_1 & 0\end{array}\right)~, &
{}~~\G^2 = \left(\begin{array}{cc}0 & -i\t_2 \\ +i\t_2 &
0\end{array}\right)~,\cr
\G^3 = \left(\begin{array}{cc} 0 & +\t_3 \\ -\t_3 & 0 \end{array}\right)~, &
{}~~\G^4 = \left(\begin{array}{cc} 0 & +iI_2 \\ +iI_2 & 0 \end{array}\right)~,}
\eqno(2.13) $$
since in this representation we have
$$\G_5\equiv \G^1\G^2\G^3\G^4 = \left(\begin{array}{cc} I_2 & 0 \\ 0 & -I_2
\end{array}\right)~.\eqno(2.14) $$

Given an explicit representation of the $\G$-matrices, it is easy
to construct an explicit representation of the $\g$-matrices (2.10)
by forming the linear combinations
$$V_{\pm}\equiv {1\over\sqrt{2}}\left( \G^1 \pm i\G^2\right),~~W_{\pm} =
{1\over\sqrt{2}}\left( \G^3 \pm i\G^4\right)~.\eqno(2.15) $$
It follows that
$$V_{\pm}^2 = W_{\pm}^2 = 0,~\left\{ V_+,V_-\right\}=-\left\{ W_+,W_-\right\}
= 2,~VW + WV =0,\eqno(2.16) $$
the latter being true for any assignment of the subscripts $\pm$. Therefore,
each choice,
$$({\bf I}):~\g^{a}=\left( V_+,W_+\right),~\g^{\bb}=\left( V_-,W_-\right),~~
({\bf II}):~\g^{a}=(V_+,W_-),~\g^{\bb}=(V_-,-W_+)~,\eqno(2.17)$$
yields the corresponding representation for the $\g$-matrices. As we will see
later on in this Letter, this choice is just related to either the self-duality
$({\bf I})$ or the anti-self-duality $({\bf II})$, respectively. For
simplicity, we restrict ourselves to the self-duality and choose $({\bf I})$
for the rest of the Letter.

Given the explicit $\G$-matrix representation (2.13) with the diagonal $\G_5
\equiv \g_5$ as in eq.(2.14), it allows to introduce
the $2\times 2$ $\s$-matrices for the $\g$-matrices as
$$\g^{a} = \left(\begin{array}{cc} 0 & \s^{a} \\ \st^{a} & 0 \end{array}
\right),~\g^{\ab} = \left(\begin{array}{cc} 0 & \s^{\ab} \\ \st^{\ab} & 0
\end{array} \right). \eqno(2.18) $$
For the purpose of dealing with the explicit representation of $\g$-matrices,
 we find convenient to define a basis in the space of $2\times 2$ matrices by
introducing the following set
$$P_+ = \left( \begin{array}{cc} 1 & 0 \\ 0 & 0 \end{array}\right)~,~~
\t_+ = \left( \begin{array}{cc} 0 & 1 \\ 0 & 0 \end{array}\right)~,~~
\t_- = \left( \begin{array}{cc} 0 & 0 \\ 1 & 0 \end{array}\right)~,~~
P_- = \left( \begin{array}{cc} 0 & 0 \\ 0 & 1 \end{array}\right)~.
\eqno(2.19) $$

Comparing eqs. (2.13), (2.15), (2.17) ({\bf I}), (2.18) and (2.19), we find
$$\eqalign{
\s^{a} = \left(-i\sqrt{2}\t_{-},-\sqrt{2}P_{-}\right), &
{}~~\st^{a} = \left( +i\sqrt{2}\t_{-},-\sqrt{2}P_{+}\right)~; \cr
\s^{\ab} = \left( -i\sqrt{2}\t_{+},+\sqrt{2}P_{+}\right), & ~~
\st^{\ab} = \left(+i\sqrt{2}\t_{+},+\sqrt{2}P_{-}\right)~.}\eqno(2.20)$$

Since the index $a$ takes only {\it two} values, the structure of linearly
independent covariant products of $\g$-matrices is rather special:
$$ \g_3~,~~\S^{a\bb}~,~~\g_{\3b}~;\eqno(2.21)$$
In this equation, the $\g_3$ and $\g_{\3b}$ represent the {\it covariant}
products
$$\g^{a}\g^{b}=2i\ve^{ab}\g_3~,~~\g^{\ab}\g^{\bb}=2i\ve^{\ab\bb}\g_{\3b}
{}~,\eqno(2.22) $$
where the Levi-Civita symbols have been introduced, $\ve^{ab}=-\ve^{ba} =
-\ve_{ab}=\ve_{ba},~\ve_{12}=-1;~~\ve^{\ab\bb}=-\ve^{\bb\ab}=-\ve_{\ab\bb}=
\ve_{\bb\ab},~\ve_{\bar{1}\bar{2}}=1\,$.
The $\S^{a\bb}$ in eq.(2.21) are similar to their $(3+1)$-dimensional
counterparts and take the form
$$\S^{a\bb} = -{1\over 4}\left( \g^{a}\g^{\bb} - \g^{\bb}\g^{a} \right)~.
\eqno(2.23) $$
Those four matrices are covariantly reducible as
$$\S^{a\bb} = {1\over 2}\h^{a\bb}\S + \Hat{\S}^{a\bb}~,\eqno(2.24)$$
where we have introduced the irreducible pieces of $\S^{a\bb}$ as
$$ \S \equiv \h_{a\bb}\S^{a\bb}=\g_{\3b 3}-\g_{3\3b},~~\h_{a\bb}\Hat{\S}^{a\bb}
 =0~.\eqno(2.25) $$

The commutators involving the $\g_3,~\g_{\3b}$ and $\S^{a\bb}$ take the form
$$\[ \g_3 , \g_{\3b}\] = -\h_{a\bb}\S^{a\bb}=-\S ~,~~
\[ \g_3 , \S^{a\bb}\] =\h^{a\bb}\g_3~,~~\[ \g_{\3b},\S^{a\bb}
\] = -\h^{a\bb}\g_{\3b}~,$$
$$\[ \S^{a\bb},\S^{c\db}\] = \h^{a\db}\S^{c\bb} -
\h^{c\bb}\S^{a\db}~.\eqno(2.26)$$
This set of matrices  forms the Lie algebra which is in fact isomorphic to
$u(1,1)$, as it should. The $\left\{ \S^{a\bb}\right\}$ form the subalgebra
which is isomorphic to $u(2)$. Clearly, the  $\left\{ \g_3, \g_{\3b},
\S^{a\bb}\right\}$ are nothing but the $U(1,1)$ generators in the spinor
representation, while $\left\{ \S^{a\bb}\right\}$ represent the generators
of the $U(2)$ subgroup in the same representation. Eq.(2.24) reflects the
relevant Lie algebra decomposition $u(2)=su(2)\oplus u(1)\,$.

In the representation (2.20), the $\g_3$, $\g_{\3b}$ and $\S$ take the form
$$\g_3 = \left( \begin{array}{cc} \t_{-} & 0 \\ 0 & 0 \end{array} \right),~
\g_{\3b} = \left( \begin{array}{cc} \t_{+} & 0 \\ 0 & 0 \end{array} \right),~
\S =\left( \begin{array}{cc} \t_3 & 0 \\ 0 & 0\end{array}\right)~.\eqno(2.27)$$
Hence, the $\g_3$ and $\g_{\3b}$ are nilpotent and hermitian-conjugated to each
 other, $\g_3^{\dg} = \g_{\3b}$. Each matrix in eq.(2.27) commutes
with the $\g_5$. In addition, they satisfy the funny relations:
$$ \g_3 = \g_5 \g_3~,~~\g_{\3b} = \g_5 \g_{\3b}~,~~\S =\g_5\S ~.\eqno(2.28) $$

The $\Hat{\S}^{a\bb}$ is also commuting with the $\g_5\,$, but satisfies the
relation
$$ \g_5\Hat{\S}^{a\bb}=-\Hat{\S}^{a\bb}~.\eqno(2.29)$$

The chiral projection operators can be rewritten as
$${1\over 2}\left( 1 + \g_5 \right) = \left( \g_3 + \g_{\3b}\right)^2 =
\g_3\g_{\3b} + \g_{\3b}\g_3  ~,~~
{1\over 2}\left( 1-\g_5\right)=1-\g_{\3b}\g_3 -\g_3 \g_{\3b}~.\eqno(2.30) $$

It is interesting that there are more {\it hermitian projectors} for spinors
in $2 + 2$ dimensions. Defining $\g_{3\3b}\equiv \g_3 \g_{\3b}~,~~\g_{\3b 3}
\equiv \g_{\3b}\g_3~$, we find
$$\g^2_{3\3b}=\g^{\dg}_{3\3b}=\g_{3\3b}~,~~\g^2_{\3b 3} = \g^{\dg}_{\3b 3} =
\g_{\3b 3}~.\eqno(2.31)$$
In the explicit representation (2.20), we have
$$\g_{\3b 3} =\g_{\3b}\g_3 = \left( \begin{array}{cc} P_{+} & 0 \\ 0 & 0
\end{array}\right),~~\g_{3\3b}=\g_3\g_{\3b} = \left( \begin{array}{cc} P_{-}
& 0 \\ 0 & 0 \end{array}\right)~.\eqno(2.32) $$

\noindent {\bf 3.~~Spinors}

Let's come back to the Dirac's equation in the real form (2.7) and see what
{\it spinors} can be introduced in $2 + 2$ dimensions. One of the key issues
in this matter is the {\it charge conjugation matrix} we are going to discuss
in this section (see also ref.[10] as for a general situation in $s + t$
dimensions).

First, we notice that the $\G$-matrices can be chosen in the way such that
$$(\G^1)^{\dg} = +\G^1,~(\G^2)^{\dg} = +\G^2,~
(\G^3)^{\dg} =-\G^3, ~(\G^4)^{\dg} =-\G^4~.\eqno(3.1) $$

In particular, the explicit representations (2.11) and (2.13) do satisfy
eq.(3.1). It follows that
$$ (\G^j)^{\dg} = -\G_0\G^j\G_0^{-1}~,\eqno(3.2) $$
where the $(2,2)$ analogue of the $(3 + 1)$-dimensional $\G_0$-matrix has been
defined as
$$ \G_0\equiv \G^1\G^2,~\G_0^{-1}=\G_0^{\dg}=-\G_0,~\G^2_0=-1~.\eqno(3.3)$$

To introduce the charge conjugation matrix $C$ in the usual way, one notices
that both $\pm(\G^j)^*$ and $\pm(\G^j)^T$ separately form equivalent
representations of the Clifford algebra (2.8). Therefore, there exist
invertible matrices $B$ and $C$ for which
$$ (\G^j)^* = \h B\G^j B^{-1}~,~~(\G^j)^T = -\h C \G^j C^{-1}~,\eqno(3.4) $$
where the sign $\h,~\h=\pm 1$, has been introduced. The correlation of the
overall signs in these two equations appears to be needed for their
consistency with eq.(3.2). It also yields
$$ C = B^T \G_0~,\eqno(3.5) $$
modulo a sign factor which is irrelevant here.
It is not difficult to show that the matrix $B$ is unitary and satisfies [10]
$$ B^* B = 1~.\eqno(3.6) $$

The charge conjugation matrix $C$ has the properties
$$ C^T = -C~,~~C^{\dg}C=1~,\eqno(3.7) $$
which follow just from the definitions above. It is optional to take
$C^* =-C$.

In the Majorana representation (2.11), the matrix
 $B$ can be chosen to be $1$ while $\h=-1$. In that representation the
``Lorentz'' generators $\S^{ij}={1\over 2}\left( \G^i\G^j - \G^j\G^i\right)$
are real. Dividing them into chiral pieces $\S_{\pm}^{ij}={1\over 2}\left( 1
\pm \G_5\right)\S^{ij}$ yields two commuting $sl(2,{\bf R})$ algebras generated
by $\S_+^{ij}$ and $\S_-^{ij}$ respectively. This notice illustrates the
well-known isomorphism [10]
$$\Bar{SO(2,2)}\cong SL(2,{\bf R})\otimes SL(2,{\bf R})~,\eqno(3.8)$$
where $\Bar{SO(2,2)}$ is the covering group of $SO(2,2)$.

In the explicit representation (2.13) we have
$$(\G^1)^*=-\G^1, ~(\G^2)^* = +\G^2,~(\G^3)^* = + \G^3, ~(\G^4)^* =-\G^4.~.
\eqno(3.9) $$
It follows that
$${\bf \h=-1:} ~B=\G^3\G^2,~C=\G^1\G^3~;~~{\bf \h=+1:}~B=\G^1\G^4,~C=\G^2\G^4~,
\eqno(3.10) $$
or, more explicitly,
$${\bf \h=-1:} ~C=\left(\begin{array}{cc} +\t_2 & 0\\ 0 & +\t_2\end{array}
\right);~~{\bf \h=+1:} ~C=\left(\begin{array}{cc} +\t_2 & 0 \\ 0 & -\t_2
\end{array}\right)~.\eqno(3.11) $$
In particular, $C^* =-C$ and $\st^{\ula}=\t_2\left( \s^{\ula}\right)^T
\t_2$ (at $\h=-1$) in this representation. The $\st^{\ula}$ and $\s^{\ula}$
are in fact related by the charge conjugation matrix in any representation of
the gamma matrices.

We are now in a position to discuss the simplest spinor representations of the
lowest dimension in space-time with the $(2,2)$ signature.
The Dirac equation (2.7) defines the {\it Dirac} spinor $\J$ whose
transformation properties follow from requiring the covariance of the Dirac
equation. The Dirac spinor has 4 complex components and this representation is
clearly reducible because  of the existence of the chiral projection operators.
The chiral projectors $P_{\pm}={1\over 2}\left(1\pm\G_5\right)$ can be used to
 define the {\it chiral} or {\it Weyl} spinors, $\j_{\pm}=\pm\G_5\j_{\pm}$. In
 the convenient representation (2.13), we have
$$ \J_{\ul{\a}} =\left( \begin{array}{c} \j_{\a} \\ \Tilde{\j}_{\dt{\a}}
\end{array}\right)~,\eqno(3.12) $$
where each Weyl spinor, $\j$ or $\Tilde{\j}$, has 2 complex components,
$\a=1,2;~\dt{\a}=\dt{1},\dt{2};~{\ul{\a}}=(\a,\dt{\a})$. As in
$(3 + 1)$-dimensions, the antisymmetric
tensors, $C^{\a\b},~C_{\a\b}$ and $C^{\dt{\a}\dt{\b}},~C_{\dt{\a}\dt{\b}}$,
defined by chiral pieces of the charge conjugation
matrix in eq.(3.11), can be used to raise and lower the chiral spinor indices
$\a$ and $\dt{\a}$. It is the specialty of the $(2,2)$ space-time
that the projections $P_{\pm}\G^i$ {\it separately} transform under the
$B$ conjugation of eq.(3.4).

The complex conjugation of the Dirac equation (2.7) yields
$$\left[ i\G^j \pa_j + (-\h)m\right]\left( B^{-1}\J^* \right)=0~.
\eqno(3.13) $$
Given $(-\h)m= m$, this is the Dirac equation for $\J^c\equiv B^{-1}\J^*$
. Since $B^* B=1$, it is consistent to equate
$$ \J^* =B\J ~~{\rm or}~~\J^c = \J~.\eqno(3.14) $$
The $\J^c$ is known as the {\it Majorana-conjugated} spinor, while eq.(3.14) is
known as the {\it Majorana} condition. In the massive case, this is only
possible if $\h=-1$. In the massless case there is another option, $\h=1$,
which allows to define the {\it pseudo-Majorana} spinors as those satisfying
eq.(3.14). Introducing the $(2,2)$ analogue of the Dirac conjugation as
$$ \Bar{\J} = \J^{\dagger} \G_0 , \eqno(3.15) $$
we can rewrite the Majorana condition (3.14) to the form which is familiar from
$(3 + 1)$-dimensions:
$$\J^c = C\Bar{\J}^T~,\eqno(3.16) $$
provided the representation of $\G$-matrices, in which $C^* =-C$, is used.
In the representation (2.13) the latter takes place, while the Majorana
condition for the spinor (3.12) yields $\t_1\j=\j^*$ and $\t_1\Tilde{\j}=
\Tilde{\j}^*$. Clearly, this distinguishes the $(2,2)$ case from its $(3,1)$
counterpart where we had $\j^* =\Tilde{\j}$ instead.

It is now obvious that we can introduce {\it Majorana-Weyl} (MW) spinors in
$2 + 2$ dimensions without adding additional isospin indices. The chiral parts
 of the Majorana spinor just represent the MW-spinors. In other words, the two
constraints provided by the chirality and Majorana conditions can be
simultaneously and consistently imposed on a spinor in $2 + 2$ dimensions! In
the Majorana representation (2.11) with pure imaginary $\G^j$, there exist {\it
  real} (Majorana) spinors, which satisfy the real Dirac equation. Since in the
 Majorana representation the $\G_5$ is off-diagonal but {\it real}, the
chirality condition makes perfect sense and leads to a 2-component {\it real
chiral} spinor which is nothing but the MW-spinor. In the representation
(2.13), a 4-component MW-spinor $\J$ takes the form
$$\J_{\ul{\a}} = \left(\begin{array}{c} \j_{\a} \\ 0\end{array}\right),~~~
\j_{\a}=\left(\begin{array}{c} \j \\ \j^*\end{array}\right)~,\eqno(3.17)$$
where the $\j$ has just one complex component. Finally, from the viewpoint of
the isomorphism (3.8), the MW spinor just realizes the fundamental
(2-dimensional and real) representation of one of the $SL\left(2,{\bf R}
\right)$ factors in eq.(3.8).

It is worth to mention here that the {\it little} group
 for the $U(1,1)$ ``Lorentz'' group, is just $U(1)$.
The $U(1)$ is an abelian group whose irreducible representations (irreps) are
one-dimensional. To have a non-trivial spin under the $U(1)$ little group, the
$U(1,1)$ irrep should be {\it massive}. The {\it massless} $U(1,1)$ irreps
consist only of scalars. The meaning of the massless scalars in reference to
 the massless states with non-trivial spins has already been explained by
Ooguri and Vafa [4]. Those scalars appear to be the potentials for the {\it
self-dual} fields which may have non-trivial spin but only one physical
degree of freedom [4].
\vglue.2in

\noindent {\bf 4.~Supersymmetric self-dual Yang-Mills theory}

The natural field system, which comprises all of the remarkable
features mentioned above, is the $N=1$ supersymmetric {\it self-dual
Yang-Mills} (SDYM) theory in $2 + 2$ dimensions \footnote{See ref.[11] for
earlier considerations of the supersymmetric SDYM theory. The self-duality
 makes sense in a space-time with the $(2,2)$ or $(4,0)$ signature, but
not with the $(3,1)$ one.}.

The $N=1$ super-Yang-Mills (SYM) theory in four space-time dimensions is
described off-shell in terms of the $N=1$ (non-abelian) vector multiplet which
contains (in the Wess-Zumino gauge) a gauge (Lie algebra-valued) vector
field $A_i$, one Majorana spinor $\L$ and an auxiliary scalar $D$, all in
the adjoint representation of the gauge group [12]. In the associated $N=1$
superspace, the off-shell $N=1$ SYM theory is well-known to be described by a
single Majorana spinor superfield strength $W$ comprising the component fields.
  All the other $N=1$ SYM superfield strengths are expressed in terms of the
$W$, and the leading component of this superfield in the usual $\q$-expansion
over anticommuting superspace coordinates just gives the spinor compoment,
 $\left.\L = W\right|$.

The celebrated self-duality condition [6] on the Yang-Mills field strength
in four space-time dimensions with the $(2,2)$ signature is given by
$$ F_{ij} = {1\over 2}\ve_{ij}^{~~kl}F_{kl} \eqno(4.1) $$
in the {\it real notation}, where the totally antisymmetric Levi-Civita
symbol $\ve^{ijkl}$ with unit weight has been introduced. Given the {\it
complex} notation, eq.(4.1) is equivalent to the {\it two} conditions [5]
$$ F_{ab} = F_{\ab\bb}=0~,\eqno(4.2a) $$
$$ \h^{a\bb}F_{a\bb}=0~.\eqno(4.2b) $$

Any real $(2,2)$-dimensional vector $V_j=\left( \cv_1,\cv_2,\cv_3,\cv_4\right)$
can be represented by a complex pair $V_{a}=\left(V_1,V_2\right)$ in the
complex notation, where $V_1=\cv_1 + i\cv_2$ and $V_2=\cv_3 + i\cv_4$. Given
that, we have $V_{\ab}=\left( V_{\bar{1}},V_{\bar{2}}\right)=\left( \Bar{V}_1,
\Bar{V}_2\right)$, respectively. The $\s$-matrices of eq.(2.20) can be used to
 convert a vector index into a pair of spinor indices and vice versa.

Given the Majorana spinor $W$ in the form (3.12),
and the complex notation, the SYM equations of motion in the superspace read:
$$i(\s^{\ula})_{\a}{}^{\dt{\a}}\nabla_{\ula}\Tilde{w}_{\dt{\a}}=i(\st^{\ula}
)_{\dt{\a}}{}^{\a}\nabla_{\ula}w_{\a}=0~,\eqno(4.3a)$$
$$\nabla^{\ulb}F^{I}_{\ula\ulb}=-if^{IJK}\h_{\ula\ulc}\left[ w^J_{\b}
C^{\b\a}(\s^{\ulc})_{\a}{}^{\dt{\a}}\Tilde{w}^K_{\dt{\a}}\right]~,\eqno(4.3b)$$
where the gauge group structure constants $f^{IJK}$  and the gauge-covariant
derivatives $\nabla_{\ula}$ have been introduced.

The first self-duality condition on the YM field strength in eq.(4.2a) can be
interpreted as an {\it integrability condition} for the existence of the {\it
holomorphic} and {\it anti-holomorphic} spinors satisfying the equations
$$\nabla_a \J = 0~,~~\nabla_{\ab}\O  = 0~,\eqno(4.4)$$
respectively. The YM fields satisfying eq.(4.2a) can be referred
to as the ``{\it hermitian}'' gauge configurations. This equation can easily
be solved in terms of the prepotential {\it scalar} \footnote{See the
discussion at the end of sect. 3.} fields $J$ and $\Bar{J}$ as [13]
$$ A_a =J^{-1}\pa_{a}J~,~~A_{\ab}=\Bar{J}^{-1}\pa_{\ab}\Bar{J}~.\eqno(4.5)$$

The ``hermiticity'' condition (4.2a) for the Yang-Mills gauge fields has the
counterpart in the theory of gravity. This counterpart is known as the
``K\"{a}hlerness'' condition which means that the base manifold (curved
space-time) should be a hermitian manifold while the endowed space-time
connection is to be consistent with the complex structure [14].

Eq.(4.2b) puts the Yang-Mills theory on-shell, and it is analogous to the
Ricci-flatness condition in the theory of gravity. As is well-known in four
dimensions [15], a K\"{a}hlerian and Ricci-flat manifold has a self-dual
curvature tensor and vice versa. Quite similarly, eqs. (4.1) and (4.2) for the
Yang-Mills system are equivalent.

We are now ready to address the self-duality conditions in the
supersymmetric case by transforming the relevant constraints (4.2) from the
Yang-Mills field strength $F$ to its spinor superpartner $\L$. It is usually
the case in superspace that the origin of constraints on bosonic fields can be
tracked back to constraints on some fermionic fields in lower-dimensional
sector. In our case we can easily suspect that the supersymmetric SDYM
condition is to be on the $W$-superfield.
To get the supersymmetric SDYM constraint, first we notice that
the supersymmetry variation of $\L$ is just proportional to $F$:
$$ \d\L = \S^{ij}F_{ij}\ve ~,\eqno(4.6) $$
 where the Majorana spinor parameter $\ve$ of rigid $N=1$
supersymmetry and the ``Lorentz'' generators $\S^{ij}$ in the spinor
representation have been introduced (in the real notation).

In the complex notation, eq.(4.6) can be rewritten as
$$\d\L=\left( 2i\g_3\ve^{ab}F_{ab} + 2i\g_{\3b}\ve^{\ab\bb}F_{\ab\bb}+{1\over
2}\S\h^{a\bb}F_{a\bb} + \Hat{\S}^{a\bb}F_{a\bb}\right)\ve ~.\eqno(4.7) $$

Now we can separate the vanishing SDYM field strengths from the
non-vanishing ones simply by taking the chiral projections of eq.(4.7), because
of eqs. (2.28) and (2.29). This observation immediately gives rise to the
supersymmetric SDYM constraint in the form of the {\it Majorana-Weyl}
 condition on the SYM spinor superfield strength $W$:
$${1\over 2}\left( 1 + \g_5\right)W_{\rm SD} =0~.\eqno(4.8)$$

In order to check this conditon, first we notice the consistency of the
self-duality equations (4.2) with the equations of motion in eq.(4.3) for the
SYM gauge field strength $F$ just because the gaugino source term on the RHS
of the YM field equation is now {\it vanishing}, subject to the supersymmetric
 SDYM condition (4.8). Second, we can also confirm that the gaugino field
equations in eq.(4.3) come out of the spinorial derivatives \footnote{See
ref.[5] for more details.} $\nabla_{\ul{\a}}F_{ab}^I=0$ and $\nabla_{\ul{\a}}
\left(\h^{a\bb}F_{a\bb}^I\right)=0\,$. Third, the analysis of the
superspace SYM Bianchi identities subject to the constraint (4.8), whose
details we skip here, yields the outcome that the chiral part $w$ of the
Majorana superfield $W$ must be simultaneously holomorphic and
anti-holomorphic, $\nabla_a w = \nabla_{\ab}w=0$, which implies $w=0$. In
summary, we have found that the MW condition (4.8) generates nothing else than
the SDYM conditions (4.2).

Given the second choice $({\bf II})$ in eq.(2.17) to represent the
$\g$-matrices, the similar analysis would give rise to the anti-self-dual (ASD)
 SYM theory, characterized by the constraint
$${1\over 2}\left( 1 - \g_5\right)W_{\rm ASD}=0~.\eqno(4.9)$$

Before concluding this section, we mention that
the {\it extended} supersymmetric SDYM models in $2+2$ dimensions can also be
introduced. To give an example, consider the $N=2$ supersymmetric YM multiplet
defined by the $N=2$ superspace constraints \footnote{
Here we are using basically the same notation for the spinors, namely
all the {\it undotted} (or {\it dotted}) spinors have positive (or
negative) chiralities.  The $N=2$ spinor charges now carry the additional
indices $~i,\,j,\,\cdots= 1,\,2$~ for the 2-dimensional representation of
$Sp(1)$ group.  Both of $~S^I$~ and $~T^I$~ are {\it real} scalars, carrying
the adjoint index $~I$~ of the gauge group.} ({\it cf.} [16])
$$\eqalign{&F_{\a i \,\ulb}{}^I = - i(\s_\ulb)_{\a\dt\b} \Tilde w^{\dt\b}{}
_i{}^I ~~,  ~~~~
F_{\dt\a i \,\ulb}{}^I = -i(\s_\ulb)_{\b\dt\a}w^\b{}_i{}^I ~~, \cr
&\nabla_{\a i} S^I = - w_{\a i} {}^I ~~,~~~~ \Tilde\nabla_{\dt\a i} S^I = 0
{}~~, \cr
&\Tilde\nabla_{\dt\a i} T^I = -\Tilde w_{\dt\a i} {}^I ~~, ~~~~
\nabla_{\a i}T^I = 0~~, \cr
&F_{\a i\,\b j}{}^I = 2C_{\a\b} \e_{i j} T^I ~~, ~~~~
F_{\dt\a i\,\dt\b j} {}^I = 2C_{\dt\a\dt\b} \e_{i j} S^I ~~,
{}~~~~ F_{\a i\,\dt\b j} {}^I = 0~~. \cr} \eqno(4.10) $$

The self-duality condition is now
imposed as $~S^I=0$, whereupon the component fields $~w_{\a i}{}^I$~
and $~F_{a b}{}^I,~F_{\Bar a\Bar b}{}^I$~ and $~\eta^{a\Bar b}F_{a\Bar b}
{}^I$~ vanish simultaneously.   Similarly, we can get a SDYM multiplet even
with $N=4$ supersymmetry, whose details we will give in a future
publication [17].
\vglue.2in

\noindent{\bf 5.~~Concluding remarks}

The relevance of a four-dimensional space-time with the $(2,2)$ signature
appears to be based on the very plausible assumption that the underlying
theory of all exactly solvable (integrable) models generated by
the SDYM theory in $2 + 2$ dimensions, might be just the $N=2$ heterotic
string. Space-time supersymmetry by itself is a good motivation to introduce
the supersymmetric SDYM models and the MW spinors associated with them.
The supersymmetric SDYM model in $2 + 2$ dimensions or, maybe, its maximal
$N=4$ supersymmetric extension, have a good chance to be the master theory for
all supersymmetric integrable models in lower dimensions.

To our knowledge, the supersymmetric models in space-time with the $(2,2)$
signature have {\it never} been explicitly presented in the past, not to
mention the supersymmetric SDYM multiplets and the real scalar supermultiplet.

However, the connection, if any, between $N=2$ strings and space-time
supersymmetry is yet to be found. The spectra of $N=2$ strings comprise only
scalars [4] with no supersymmetry. It is well-known that the Green-Schwarz
superstrings, when coupled to supersymmetric Yang-Mills and/or
supergravity backgrounds, do require the supersymmetric Yang-Mills and/or
supergravity fields to satisfy their equations of motion [18]. Technically, it
 originates from the Siegel's invariance of the coupled GS action. Taking the
supersymmetric space-time background on-shell just enough for the consistency
of the theory, but the self-duality condition does {\it not} violate that
consistency [8].

To this end, we would like to mention two related results about the existence
of the {\it self-dual} supergravity (SDSG) and a {\it real
scalar} supermultiplet in $2 + 2$ dimensions! Given the curvature tensor valued
in the ``Lorentz'' algebra $u(1,1)$, the self-duality conditions take the form
$$ R_{ab}{}^{\ulc\uld} = R_{\ab\bb}{}^{\ulc\uld} = 0~,
{}~~\h^{a\bb}R_{a\bb}{}^{\ulc\uld} = 0~,\eqno(7.1) $$
which are quite similar to the SDYM counterparts in eq.(4.2). Taking into
account [9] that (i) the $N=1$ on-shell supergravity in superspace is
described by the irreducible superfield strengths $W_{\a\b\g}$  and
$\Tilde{W}_{\dt{\a}\dt{\b}\dt{\g}}$ (totally symmetric on their spinor
indices), and (ii) the leading components of those superfields are just the
gravitino fields, the natural ansatz for the $N=1$ SDSG constraint in $N=1$
superspace is given by [17]
$$W_{\a\b\g}=0~.\eqno(7.2)$$

The real scalar $N=1$ multiplet consists of a {\it real} scalar $A$, {\it
Majorana-Weyl} spinor $\Tilde{\j}$ and a {\it real} auxiliary scalar $F$. The
$N=1$ supersymmetry transformation laws (with the {\it real} parameters
$\ve_{\a}$ and $\Tilde{\ve}_{\dt{\a}}$) read:
$$\d A =-i\Tilde{\ve}_{\dt{\b}}C^{\dt{\b}\dt{\a}}\Tilde{\j}_{\dt{\a}}~,~~
\d\Tilde{\j}_{\dt{\a}}=i(\st^{\ula})_{\dt{\a}}{}^{\b}\pa_{\ula}A\ve_{\b} +
F\Tilde{\ve}_{\dt{\a}}~,~~\d F =-\ve_{\a}C^{\a\b}\left(\s^{\ula}\right)_{\b}
{}^{\dt{\g}}\pa_{\ula}\Tilde{\j}_{\dt{\g}}~.\eqno(7.3)$$
Hence, this 4-dimensional multiplet is just like a 2-dimensional scalar
multiplet! The details will be reported elsewhere [17].
\vglue.2in
\oddsidemargin=0.03in
\evensidemargin=0.01in
\hsize=6.5in
\textwidth=6.5in

\baselineskip 6 pt

\noindent {\bf References}

\def\sNP{{\sf Nucl. Phys.~}}
\def\sPL{{\sf Phys. Lett.~}}
\def\sCMP{{\sf Commun. Math. Phys.~}}
\def\sIJMP{{\sf Int. J. Mod. Phys.~}}
\def\sCQG{{\sf Class. Quantum Grav.~}}
\def\sPRL{{\sf Phys. Rev. Lett.~}}
\def\sMPL{{\sf Mod. Phys. Lett.~}}

\begin{description}
\item[{[1]}] I. Ya. Aref'eva and I. V. Volovich, \sPL {\bf 164B} (1985) 287 ;\\
A. D. Popov, \sPL {\bf 259B} (1991) 256.
\item[{[2]}] D. J. Bruth, D. B. Fairle and R. G. Yates, \sNP {\bf B108} (1976)
310 ; \\
M. Ademollo, L. Brink, A. D'Adda, R. D'Auria, E. Napolitano, S. Sciuto, E. Del
Giudice, P. Di Vecchia, S. Ferrara, F. Gliozzi, R. Musto and R. Pettorino,
\sPL {\bf 62B} (1976) 105 ; \\
M. Ademollo, L. Brink, A. D'Adda, R. D'Auria, E. Napolitano, S. Sciuto, E. Del
Giudice, P. Di Vecchia, S. Ferrara, F. Gliozzi, R. Musto, R. Pettorino and
J. H. Schwarz, \sNP {\bf B111} (1976) 77 ; \\
L. Brink and J. H. Schwarz, \sNP {\bf B121} (1977) 283
\item[{[3]}] E. S. Fradkin and A. A. Tseytlin, \sPL {\bf 106B} (1981) 63 ; \\
S. D. Mathur and S. Mukhi, \sNP {\bf B302} (1988) 130
\item[{[4]}] H. Ooguri and C. Vafa, \sMPL {\bf A5} (1990)
1389; \sNP {\bf B361} (1991) 469; {\bf B367} (1991) 83
\item[{[5]}] H. Nishino and S. J. Gates Jr.,  Maryland preprint,
UMDEPP 92--137, January 1992
\item[{[6]}] A. A. Belavin, A. M. Polyakov, A. Schwartz and Y. Tyupkin, \sPL
{\bf 59B} (1975) 85 ; \\
R. S. Ward, \sPL {\bf 61B} (1977) 81 ; \\
M. F. Atiyah and R. S. Ward, \sCMP {\bf 55} (1977) 117 ; \\
E. F. Corrigan, D. B. Fairlie, R. C. Yates and P. Goddard, \sCMP {\bf 58}
(1978) 223 ; \\
E. Witten, \sPRL {\bf 38} (1977) 121 ;\\
A. N. Leznov and M. V. Saveliev, \sCMP {\bf 74} (1980) 111 ; \\
L. Mason and G. Sparling, \sPL {\bf 137B} (1989) 29 ;\\
I. Bakas and D. A. Depireux, \sMPL {\bf A6} (1991) 399; 1561; 2351.
\item[{[7]}] R. S. Ward, {\sf Phil. Trans. Roy. Lond.} {\bf A315} (1985) 451 ;
\\
N. J. Hitchin, {\sf Proc. Lond. Math. Soc.} {\bf 55} (1987) 59 ; \\
A. A. Belavin and V. E. Zakharov, \sPL {\bf 73B} (1978) 53.
\item[{[8]}] H. Nishino, S. J. Gates Jr. and S. V. Ketov, Maryland preprint,
UMDEPP 92--171, February 1992
\item[{[9]}] S. J. Gates Jr., M. T. Grisaru, M. Rocek and W. Siegel, {\it
Superspace}, Benjamin/Cummings, Reading MA, 1983
\item[{[10]}] T. Kugo and P. D. Townsend, \sNP {\bf B221} (1983) 357
\item[{[11]}] I. V. Volovich, \sPL {\bf 123B} (1983) 329 ; \\
C. R. Gilson, I. Martin, A. Restuccia and J. C. Taylor, \sCMP {\bf 107}
(1986) 377
\item[{[12]}] J. Wess and B. Zumino, \sNP {\bf 79} (1974) 413
\item[{[13]}] Y. Brihaye, D. B. Fairlie, J. Nuyts and R. G. Yates, {\sf
Jour. Math. Phys.} {\bf 19} (1978) 2528 ; \\
K. Pohlmeyer, \sCMP {\bf 72} (1980) 37
\item[{[14]}] E. K\"{a}hler, {\sf Abh. Math. Sem. Univ. Hamburg} {\bf 9}
(1933) 173
\item[{[15]}] M. F. Atiyah, N. J. Hitchin and I. M. Singer, {\sf Proc. Roy.
Soc. Lond.} {\bf A262} (1978) 425
\item[{[16]}] R. Grimm, M. Sohnius and J. Wess, \sNP {\bf 1978} (1978) 275
 \item[{[17]}] S. J. Gates Jr., S. V. Ketov and H. Nishino, Maryland preprint,
in preparation (March 1992).
\item[{[18]}] E. Witten, \sNP {\bf B268} (1986) 79 ; \\
A. Sen, \sNP {\bf B268} (1986) 287 ; \\
C. M. Hull, \sPL {\bf 178B} (1986) 357 ;\\
M. T. Grisaru, A. van de Ven and D. Zanon, \sPL {\bf 173B} (1986) 423; \sNP
{\bf B277} (1986) 388.
\end{description}

\end{document}